\documentclass[10pt,landscape,twocolumn]{article}
\usepackage{geometry}
\usepackage{graphicx}        % standard LaTeX graphics tool
\usepackage{float}
\usepackage[labelfont=bf,figurename=Fig.]{caption}
\makeatletter

\renewcommand{\section}{\@startsection
   {section}%
   {1}%
   {5mm}%
   {6pt}%
   {6pt}%
   {\bf}}%
\renewcommand{\subsection}{\@startsection
   {subsection}%
   {2}%
   {5mm}%
   {6pt}%
   {6pt}%
   {\small\bf}}%
\renewcommand{\part}{\@startsection
   {part}%
   {0}%
   {0mm}%
   {6pt}%
   {6pt}%
   {\bf}}%
\renewcommand{\@biblabel}[1]{#1.\hfill}

\usepackage{xspace}
\usepackage{ulem}

\def\nifsx{${}^{56}$Ni\xspace}

\def\onehalf{ \hbox{${1\over 2}$} }
\def\pd{\partial}

\def\Bxout{\bgroup \markoverwith{\raise .3ex\hbox{$\Bigl\backslash$}}\ULon}
\newcommand{\mydotuline}{\bgroup \markoverwith{\lower .55ex\hbox{.}}\ULon}

\begin{document}
\begin{center}
% \large{\textbf{Радиативные ударные волны и их роль в объяснении загадки
%     Сверхмощных сверхновых}}
%

\large{\textbf{Radiative shock waves and their role in solving puzzles of Superluminous Supernovae}}

\vspace{12pt}
\textit{ S.I.~Blinnikov$^{1,2}$}
\vspace{6pt}

\small{$^{1}$ITEP, 117218, Moscow, Russia, Sergei.Blinnikov@itep.ru\\
$^{2}$Kavli IPMU (WPI), The University of Tokyo, Kashiwa, 277-8583, Japan}
\end{center}
% \twocolumn
% \begin{multicols}{2}
\baselineskip=11pt
\vspace{12pt}

\section{Introduction: Three paths, i.e. three scenarios proposed for SLSNe}

Typical supernova (or SN) explosions produce ejecta with kinetic energies $10^{51}$~ergs~$\equiv 1$~foe.
This unit of energy was introduced by H.Bethe and is an abbreviation of 10 to \textbf{fifty one ergs}.
Light emitted during the first year of supernova is only a small fraction, around 0.01~foe.
Many supernovae are discovered in last decade with peak luminosity one-two orders of magnitude higher than
for normal supernovae of known types.
They emit powerful light with energy approaching 1~foe, and sometimes even higher.
\begin{figure}[H]
\centering
\includegraphics[width=0.7\linewidth]{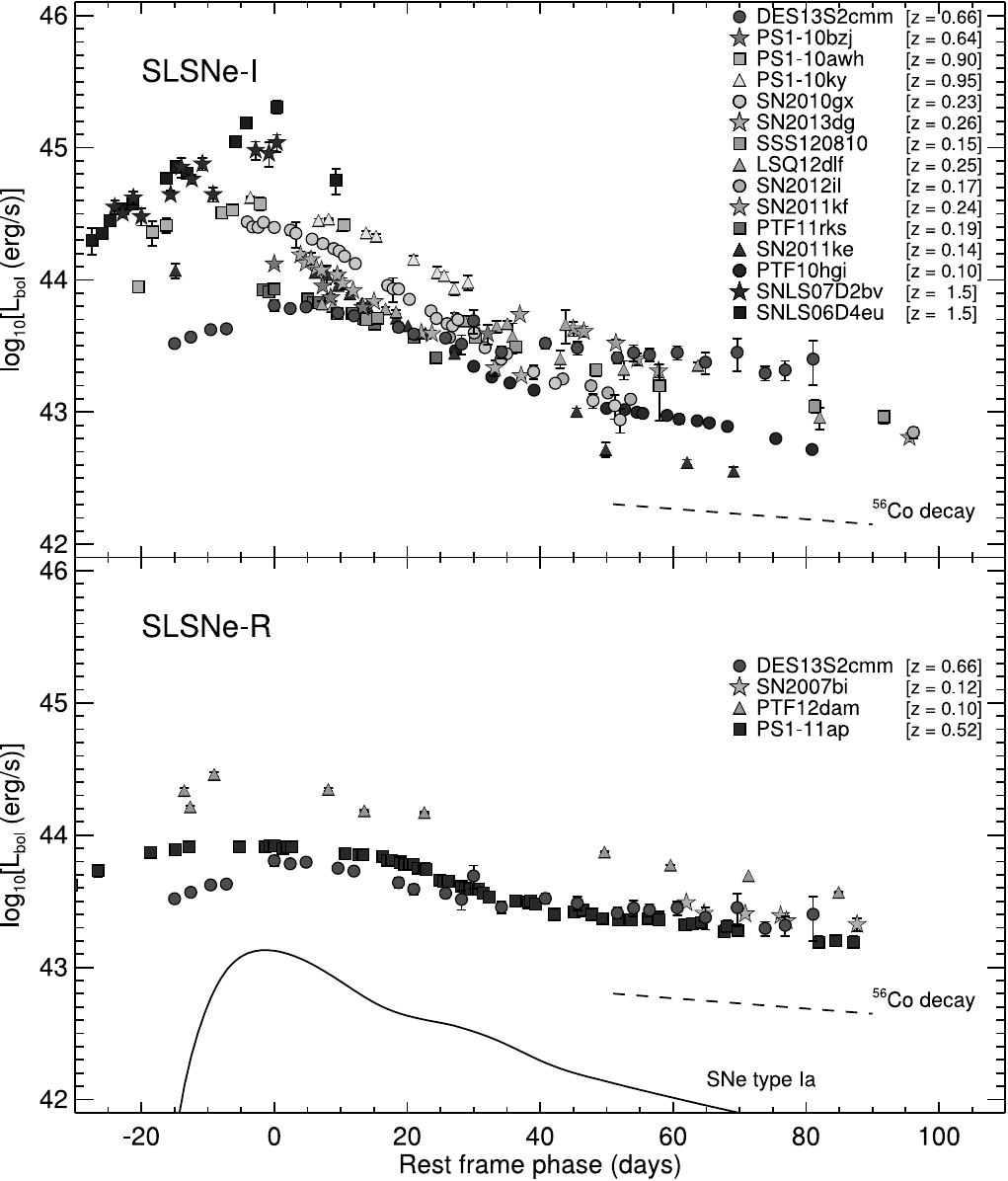}
\caption{\noindent\small  Bolometric light curves of some Superluminous Supernovae.
From paper~[1] % Papadopoulos et al. 1915
        }
\label{Papadopoulos}
\end{figure}
An example is given in Figure~\ref{Papadopoulos} together with a typical SN~Ia (used for cosmology due to their
high luminosity) for comparison.
We see that the new objects emit much more light.
They are called Superluminous supernovae: SLSNe.
This is a challenge for theory, since even normal supernovae are
not yet completely understood from the first principles.

% \begin{frame}{{Три пути для SLSNe}}

Many models are discussed in literature on SLSNe, but it seems that the most viable are the following three scenarios.

\begin{enumerate}
%  \item {\color{red} Неустойчивость при рождении пар Pair instability Supernovae, \textbf{PISN}}
 \item {Pair Instability Supernovae, \textbf{PISN}}
 \item  \textbf{``Magnetar''} pumping (taking in quotes, since observed magnetars are slowly
 rotating in Soft Gamma-ray Repeators, and here millisecond periods are needed)
%  \item  {``Магнитарная''} накачка -- нужны миллисекундные периоды
 \item {\textbf{Shock} interaction with Circumstellar Matter (CSM), e.g. Pulsational pair instability, \textbf{PPISN}}
%  \item {\color{blue} Радиативные \bfitmy{ударные волны} в окружающей плотной оболочке, {PPISN}}
\end{enumerate}

\section{Pair Instability and Supernovae}

To understand the mechanism of explosion of Pair Instability Supernovae one has to learn
a bit on stellar evolution theory.

\begin{figure}[H]
\centering
\includegraphics[width=0.7\linewidth]{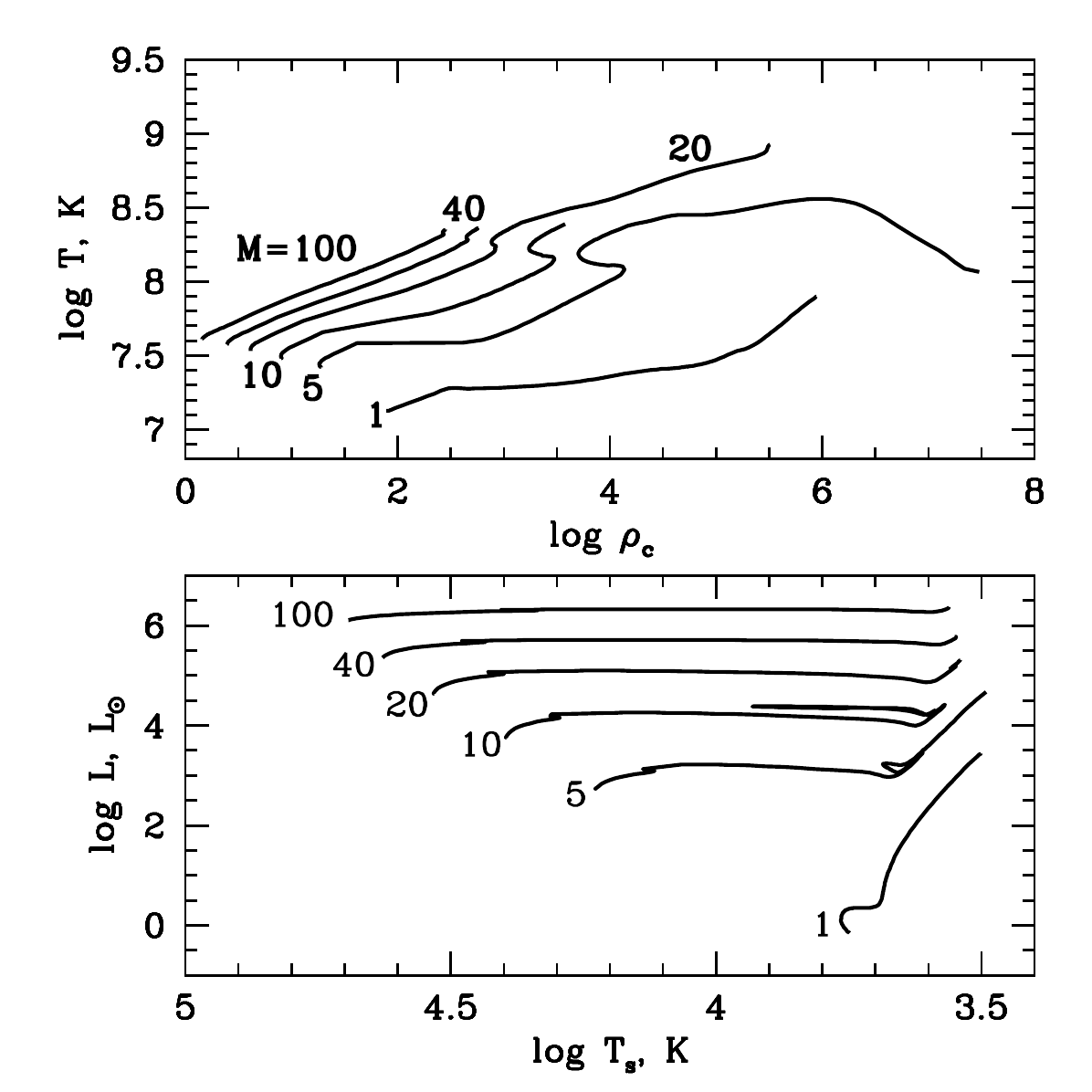}
\caption{\noindent\small Evolution of central temperature vs. central density (upper panel) and Hertzprung-Russel
diagram (lower panel) for a few models of normal stars.
Numbers near the tracks are masses of stars in units of solar mass $M_\odot$
        }
\label{setHRtrho}
\end{figure}

Figure~\ref{setHRtrho} shows the evolutionary tracks of normal stars of various masses.
One can easily notice a trend: central temperature $T_c$ grows with central density $\rho_c$ approximately
as $ T_c \propto \rho_c^{1/3} $. This law is more pronounced for more massive stars.
This relation is easily understood from the conditions of mechanical equilibrium of a star.

A very crude order-of-mag\-ni\-tude estimate
for the  attraction force of two halves of a star with mass $M$ is
$$
 F \sim {G_{\rm N}M^2\over 4R^2},
$$
where $R$ is the star radius, and $G_{\rm N}$ is the Newton's constant.
This force must be balanced by a gradient of pressure $P$ in a star which is in equilibrum.

Pressure $P$ is virtually zero on the surface, and in the center
$$
P_c={F \over S}= {F \over \pi R^2} .
$$
Omitting all coefficients of order unity, we get for the pressure and density in the center:
$$
P_c \simeq {G_{\rm N}M^2\over R^4} ,
$$
$$
\rho_c \simeq {M\over R^3} ,
$$
and we find that the equilibrium requires (in \textbf{Newtonian gravity}):
$$
{ P_c \simeq G_{\rm N}M^{2/3}\rho_c^{4/3} } .
$$
So if we have a classical ideal plasma with
$$
P={\cal R}\rho T/\mu ,
$$
where ${\cal R}$ % $= 8.314\times 10^7 {\rm erg}/({\rm g }\cdot{\rm K})$
is the universal gas constant, and $\mu$ is  mean molecular mass, we get for the central temperature
$$
 T_c \simeq {G_{\rm N}M^{2/3}\rho_c^{1/3} \mu \over {\cal R}}.
$$
Thus,
{${ T_c \propto M^{2/3}\rho_c^{1/3}}$ in non-degenerate stars}
and for a given mass
$$
{ T_c \propto \rho_c^{1/3} }
$$
The same 1/3 power is obtained for radiation-dominated massive stars (but with $ M^{1/6} $).

The condition of mechanical equilibrium,
$$
{ P_c \simeq G_{\rm N}M^{2/3}\rho_c^{4/3} } ,
$$
tells us quickly something on hydrodynamical stability of a star.
It implies that adiabatic exponent $\gamma < 4/3$ may lead to a hydrodynamical
instability.
Indeed, if  $\gamma < 4/3$ then a global compression of the star produces physical pressure which is \textbf{lower} than
needed for the equilibrium.
Then gravity wins over pressure and the star starts collapsing.

Relativistic particles, i.e. photons  lead to $ \gamma \to 4/3$.
In massive stars, with $M \sim 100 M_\odot $ we have
${ \gamma \sim 4/3 }$ due to high entropy $S$ (we have many photons per baryon).

% At low entropy ${ S \to 0}$ we have  ${ \gamma \to 4/3} $ due to { high
% Fermi energy} of {degenerate} electrons at high density $\rho$.

A radiation dominated massive star was already at the verge of the loss of the
stability, pressure was close to the law  ($P \propto \rho^{4/3}$), and now it loses it: $\gamma$ becomes less than  $4/3$
since the creation of ${ e^+e^-}$ pairs requires energy.
The momenta of particles at compression do not grow fast enough: the pairs are created with particles having low speed.
The loss of mechanical stability at $\rho_c \sim 10^4$~gcc leads to the beginning of gravitational collapse
which may be halted at $\rho_c < 10^7$~gcc by a powerful explosion, if the star has enough nuclear fuel (mostly oxygen) in the centre.
The explosion gives kinetic energy up to about 70~foe, and enough light to explain some of SLSNe due to production of huge
amount of radioactive \nifsx: up to $\sim 20 M_\odot$ (while in standard thermonuclear supernovae this amount
is 20 - 50 times smaller).

This model, which is called PISN (Pair Instability Supernova), is able to explain some of SLSN events but only
the slow ones, due to the long diffusion time of photons in ejecta overwhelmed by the iron peak elements
(products of the explosion itself and of \nifsx decay).

Similar slow light curves are produced by ``Magnetar'' pumping. There are many uncertainties in the magnetar model,
which also involves huge energy of tens foe extracted from rotational energy of a neutron star.
I do not discuss it here, and go directly to the most economical model which is especially good for
fast SLSNe.

\section{Models with radiating shock waves}

The models explaining SLSN events with the minimum energy budget
involve multiple ejections of mass in presupernova stars.
The radiative shocks produced in collisions of those shells may provide the required power of light.
This class of the models is referred to as ``interacting'' supernovae.
Nonlinear effects are important on all stages of the superluminous supernova phenomenon:
from the initial mass ejections, then at the maximum light and through
the supernova remnant with fragmentation of dense shells.

Let us obtain some estimated for important quantities.

Luminosity is expressed through effective temperature $T_{\rm eff}$ and photospheric radius $R_{\rm ph}$:
\begin{equation}
 L = 4\pi \sigma T_{\rm eff}^4 R_{\rm ph}^2
\end{equation}
for the age of supernova $t = 10$~d and typical velocity at the photosphere level
$u=10^9$~cm/s (i.e. 10 thousand km/s) we get
$R_{\rm ph} = ut \approx 10^{15}$~cm, and if a typical $T_{\rm eff} \sim 10^4$~K, then
$L \sim 10^{43}$~erg/s.

Luminosity $L$ goes down in some weeks, thus,
ordinary, non-interacting supernovae produce $\sim 10^{49}$~ergs$=0.01$~foe in photons
during the first year after explosion, while $\sim 10^{51}$~ergs$=1$~foe remain in the kinetic
energy of ejecta in ``standard'' SN explosions.

This energy is radiated much later, during millennia after the explosion
(mostly in X-rays) by the supernova remnant in the shocks produced by ejecta
in ordinary interstellar medium with the number density $\sim 1$~cm$^{-3}$.
If the density of CSM is billion times higher, then a large fraction of the kinetic
energy will be radiated away much faster, on a time scale of a year and the photons
will be much softer than X-ray, they will be emitted mostly in visible or ultraviolet
range.

We may have the same typical $T_{\rm eff}\sim 10^4$~K, while
$R_{\rm ph} \sim  10^{16}$~cm
is much larger and the luminosity goes up approaching $L \sim 10^{45}$~erg/s for some
period of time.
Thus a superluminous supernova (SLSN) can be produced for the energy of explosion
on the standard scale of 1~foe~$\sim 10^{51}$~ergs, but now a major fraction
of this energy is lost during the first year.

If we have a blob of matter with mass $m_1$ and momentum $\mathbf{p_1}$
its kinetic energy is
\begin{equation}
   E_1 = { \mathbf{p_1}^2 \over 2 m_1} .
\end{equation}
If it is colliding with another blob with mass $m_0$ and zero momentum we get for
the final energy of two merged blobs in a {fully inelastic collision}
\begin{equation}
   E_2 = { \mathbf{p_1}^2 \over 2 (m_1+m_0)} .
\end{equation}
The momentum is conserved, but the energy in amount $E_1-E_2$ is  radiated away, since $E_2 < E_1$.
If $m_0 \ll m_1$ only a tiny fraction of $E_1$ is radiated, but if  $m_0 \gg m_1$, then
$E_2 \ll E_1$ and almost all initial $E_1$ is radiated away.

This means that collisions of low mass and fast moving ejecta with heavy (dense)
slowly moving blobs of CSM are efficient in producing many photons.
Of course one should remember that in this case the momentum of the two merged
blobs may be different from the initial $\mathbf{p_1}$ if we have a directed flux of
newborn photons which carry some net momentum away.

There is no much sense in evaluating this effect using the order-of-magnitude estimates
because the details of the production of photons may be complicated.
The degree of ``inelasticity'' of the collision depend on the pattern of
hydrodynamic flow
and on the properties of emission/absorption of the plasma, e.g. on its composition.
Anyway, those details and conservation of momenta and energy must be taken into account
in full radiation hydrodynamic simulations.

Now let us find the temperature behind the shock front.
Again on the level of simple estimates for the pressure behind the shock front $P_s$
we have
\begin{equation}
 P_s \sim \rho_0 D^2 = n_0 m_i D^2
 \label{Pshock}
\end{equation}
if the density upstream the front is $\rho_0$, and $D$ is the velocity of the front.
The density $\rho=n m_i$ with $n$ number density and $m_i$ averaged mass of ions.
The estimate (\ref{Pshock}) follows from the momentum conservation: the momentum flux is
$P+\rho u^2$ for the
flow having velocity $u$, and $P$ is negligible ahead of the front where the matter
is cold.
More accurate expression for $P_s$ is easily derived from the laws of conservation.

The estimate (\ref{Pshock}) gives for
a non-relativistic plasma with pressure $P=n k_B T$:
\begin{equation}
  k_B T_s \sim {m_i D^2}
  \label{kTshock}
\end{equation}
which suggests very high temperatures, in keV range and higher for shock velocities
larger that a thousand km/s.

Now I will derive ``exact'' coefficients in (\ref{kTshock}).

Let us use standard notations for density $\rho$, velocity $u$, pressure $P$, thermodynamic energy $E$,
and define
$$
U_1=\rho ,
$$
density of momentum
$$
U_2=\rho u ,
$$
total energy density
$$
U_3=E+{\rho u^2 \over 2} .
$$

We also define fluxes of mass,
$
F_1=\rho u ,
$

\noindent of momentum,
$
F_2=\rho u^2 + P ,
$

\noindent
and of energy
$
F_3=(E+{\rho u^2/2}+ P)u ,
$
and we have a general law of conservation:
$$
  {\partial \vec U \over \partial t} = - {\partial \vec F \over \partial x} .
$$

In a stationary case, i.e. ${\pd \vec U / \pd t} = 0$, we get
$\vec F = {\rm const} $.
Introduce
$$
 j\equiv \rho u = {\rm const}, \qquad  V \equiv {1 \over \rho} \; .
$$
From $F_2=\rho u^2 + P= j^2 V = {\rm const}$ we obtain:
$$
  j^2 V_0 + P_0 = j^2 V_s + P_s  \to P_s=P_0+j^2(V_0-V_s) \; ,
$$
Subsript ``0'' for $\rho, \; V, \; u, \; P, \; E$ denotes the initial values upstream
(ahead of the shock front), while ``s'' corresponds to the values downstream, in the
shocked matter.
It is most convenient to work in the reference frame where the front is at rest, then
the speed of the shock $D$ is just $u_0$, because by definition it is measured relative
the unshocked matter.

Now  $F_3= {\rm const}$ gives:
$$
  (E_0+ \onehalf j^2 V_0 + P_0)u_0 = (E_s+ \onehalf j^2 V_s + P_s) u_s
$$
If we replace here $u_i \to jV_i$, we get:
$$
  (E_0+ \onehalf j^2 V_0 + P_0)j V_0 = (E_s+ \onehalf j^2 V_s + P_s)j V_s \; .
$$
From here
$$
  E_0V_0+ \onehalf j^2 V_0^2 + P_0V_0 = E_sV_s+ \onehalf j^2 V_s^2 + P_sV_s \; ,
$$
and
$$
  (E_0+ P_0)V_0 + \onehalf j^2 (V_0^2-V_s^2) = (E_s+ P_s) V_s \; .
$$

But
$(V_0^2-V_s^2)=(V_0-V_s) (V_0+V_s) $ and $P_s=P_0+j^2(V_0-V_s)$ obtained above implies $V_0-V_s=(P_s-P_0)/j^2$, so
% $$
%   (E_0+ P_0)V_0 + \onehalf \mydotuline{j^2} {(P_s-P_0) \over \mydotuline{j^2} }(V_0+V_s) = (E_s+ P_s) V_s \; ,
% $$
% and
$j^2$ cancels in numerator and denominator:
$$
  (E_0+ P_0)V_0 + \onehalf \Bxout{j^2} {(P_s-P_0) \over \Bxout{j^2} }(V_0+V_s) = (E_s+ P_s) V_s \; .
$$
Thus,
$$
  \left(E_0+ {P_0+P_s \over 2} \right)V_0 = \left(E_s+ {P_0+P_s \over 2} \right) V_s \; ,
$$
and we obtain a general formula for the compression in the flow (e.g. on a shock
front):
$$
  {V_s \over V_0} = {2E_0+P_0+P_s \over 2E_s+P_0+P_s } \; .
$$

An equation of state $E=E(P,V)$, or $P=P(E,V)$ gives the shock adiabat.
% E.g. for $P=(\gamma-1)E$ we find
% $$
%   {V_s \over V_0} =
%   {(\gamma+1)P_0+(\gamma-1)P_s \over (\gamma+1)P_s+(\gamma-1)P_0} \; .
% $$
For a general equation of state in a strong shock ($P_s \gg P_0, \; E_s \gg E_0$)
which is most important in supernova envelopes,
$$
  {V_s \over V_0} = {2E_0/(P_0+P_s) +1 \over 2E_s/(P_0+P_s) +1 } \approx
{1 \over 2E_s/P_s +1 }\; ,
$$
or
$$
  {\rho_s \over \rho_0}={V_0 \over V_s} \approx 1 + {2E_s \over P_s}\; ,
$$
in general case, and
$$
 {\rho_s \over \rho_0}={V_0 \over V_s} \approx 1 + {2\over \gamma -1} = {\gamma+1 \over \gamma -1} \; ,
$$
for the case of $\gamma={\rm const}$ equation of state.

Let $P=(\gamma-1)E_{\rm tr}$, where $E_{\rm tr}$ is translational internal energy,
i.e. kinetic energy of particles in plasma,
and let $E=E_{\rm tr}+Q$, where $Q$ is, e.g., ionization
potential energy.
Then in a strong shock
$$
 {\rho_s \over \rho_0}={V_0 \over V_s} \approx 1 + {2E_{2{\rm tr}}+2Q \over P_s}
   = 1 + {2\over \gamma -1} + {2Q \over (\gamma -1)E_{2{\rm tr}}} \; ,
$$
that is
$$
 {\rho_s \over \rho_0}={V_0 \over V_s} \approx {\gamma+1 \over \gamma -1}
+ {2Q \over (\gamma -1)E_{2{\rm tr}}} \; .
$$
For $\gamma=5/3$ this gives
$$
 {\rho_s \over \rho_0}={V_0 \over V_s} \approx 4 + {3Q \over E_{2{\rm tr}}} \; ,
$$
--- formula (3.71) in the famous book on shocks written by Zeldovich and Raizer.

We found from conservation of momentum ($F_2={\rm const}$)
that  $P_s=P_0+j^2(V_0-V_s)$, i.e.
$$
        j^2={P_s-P_0 \over V_0-V_s} \approx {P_s \over V_0-V_s} =
   {P_s \over V_0[1-(\gamma -1)/ (\gamma+1)]} = {P_s(\gamma+1)\over 2V_0} \; ,
$$
--- this is valid for a strong shock, constant $\gamma$ and small $Q$.
Hence,
$$
%    \rho_0^\Bxout{2} u_0^2 = {P_s(\gamma+1)\Bxout{\rho_0}\over 2} \; ,
   \rho_0^{2} u_0^2 = {P_s(\gamma+1){\rho_0}\over 2} \; ,
$$
that is
\begin{equation}
    P_s = {2 \over \gamma +1}  \rho_0  u_0^2 .
  \label{P2shock}
\end{equation}
Note that $\gamma$ here must be taken for the gas behind the strong shock since
the pressure $P_0$ is negligible and its equation of state is irrelevant.

For a non-relativistic plasma with pressure $P=\mathcal{R}\rho T /\mu$ we get from
(\ref{P2shock})
$$
   \rho_0 u_0^2 = {(\gamma+1) \mathcal{R}\rho_s T_s \over 2 \mu} \; ,
$$
so
$$
    u_0^2 = {(\gamma+1) \mathcal{R}\rho_s T_s \over 2 \rho_0 \mu} =
{(\gamma+1)^2 \mathcal{R}T_s \over 2 (\gamma-1) \mu}\; .
$$
The postshock temperature $T_s$  for the strong shock, constant $\gamma$ and small $Q$
is (from the last equation)
$$
    T_s = {2 (\gamma-1) u_0^2 \mu \over (\gamma+1)^2 \mathcal{R} } \; .
$$
For $\gamma=5/3$ we get
\begin{equation}
   T_s = { 3 u_0^2 \mu \over 16 \mathcal{R} } \; .
    \label{TSgam}
\end{equation}

If we put here $D_8=u_0/10^8$cm/s, then $D_8$ is the shock speed in thousand km/s
and we get
\begin{equation}
   T_s(\mbox{K})=2.25\times 10^7  \mu D_8^2
\label{TSkelvin}
\end{equation}
in Kelvins or
\begin{equation}
   T_s(\mbox{keV})=1.94 \mu D_8^2
\label{TSkev}
\end{equation}
in keV. Here $\mu=A/(1+Z)$ for plasma (since $n=n_{\rm baryon}/\mu=
n_{\rm ion}A/\mu=n_{\rm ion}+n_e=n_{\rm ion}+Zn_{\rm ion}$).
Note that a typical value for  $D$ in SNe is about 10 thousand km/s,
so $T$ will be of order $10^9$~K or hundreds keV.

Since $\mathcal{R} \approx k_B/m_p$ where $m_p$ is proton mass, we have
\begin{equation}
    k_B T_s \sim m_p D_s^2 .
\label{TSkin}
\end{equation}
This estimate is the same as  used in Eq.(\ref{kTshock}) if we put
$m_i = m_p$.

In supernova envelopes these numbers are misleading!

In reality plasma in supernova conditions is at least partly relativistic:
we have a huge number of photons
with $P = aT^4/3$, and so $T_s$ is appreciably lower due to high heat capacity
of photon gas.
See Equations~(\ref{Tradshock}, \ref{TKradshock})  below, they
show that, with account of radiation
for $D$ of order of a thousand km/s and for $\rho\sim 10^{-12}$~gcc, we have
$T_s=4.3 \times 10^4$~K, well below X-ray range of temperatures, but high enough
to support high $L$ for a long time at large $R$.
% }

Using  $\gamma={\rm const}$ is a favourite approximation in many papers and simulations
in astrophysics, but in supernovae it is a very bad one, and almost irrelevant.
The value of $\gamma$ varies due to ionization/excitation of atoms, and changes strongly
on the shock front when it goes through the cold layers and heats the plasma so strongly
that radiation pressure dominates downstream behind the front.
In that case (which is quite general for supernova shock breakout) the formulas
(\ref{TSkelvin},\ref{TSkev}) are not applicable and misleading.
The equations for mass, momentum and energy conservation are more complicated for
radiative shock waves, when one has to account for the transfer of
momentum and energy of photons.
Nevertheless, there are two important limiting cases for strong shocks with radiation
when simple expressions can be derived.

In the first case we may have relatively cold gas upstream with $P_0 \ll P_s $
in the strong shock, and the gas downstream is opaque with the pressure dominated
by radiation.

Due to a high heat capacity of photon gas, the temperature behind the front
is orders of magnitude lower than in (\ref{TSkelvin}),(\ref{TSkev}).

Let us put radiation pressure for $P_s$ into (\ref{P2shock}), we get
\begin{equation}
    {aT_s^4 \over 3} = {2 \over \gamma +1}  \rho_0  u_0^2 .
  \label{Pradshock}
\end{equation}
We have $\gamma=4/3$ for the radiation dominated gas, and, substituting $u_0=D$,
we obtain
\begin{equation}
    T_s =  \left( {18 \over 7 a}  \rho_0  D^2 \right)^{1/4} .
  \label{Tradshock}
\end{equation}
That is
\begin{equation}
    T_s (K) =  4.3\times 10^4 \rho_{-12}^{1/4}  D_8^{1/2} ,
  \label{TKradshock}
\end{equation}
if we normalize density for $\rho=10^{-12}$~gcc and take $D$ in units of
thousand km/s.
The temperature in reality is much less than in (\ref{TSkelvin}).

The second important case takes place when the radiation is not trapped,
its pressure and momentum
may be neglected, but when it is very efficient in heat transport.
Now the energy is not conserved, and the energy flux $F_3$ is not constant any more.
Instead of this we may  have the constancy of temperature ahead and behind the
front.
Mass and momentum conservation give as before:
\begin{equation}
P_s=P_0+j^2(V_0-V_s) .
 \label{Pmomconsv}
\end{equation}
Now both upstream and downstream, the pressure is
$P=\mathcal{R}\rho T /\mu$ with the same $T$, so the strong shock condition,
$P_s \gg P_0$  means not a high $T$ behind the front, but $\rho_s \gg \rho_0$,
and $P_s \approx \rho_0 u_0^2$, which we get from (\ref{Pmomconsv}), gives
\begin{equation}
{\rho_s \over \rho_0} = {\mu D^2 \over \mathcal{R} T } .
 \label{Tisoshock}
\end{equation}

The isothermal $T$ here is much less than the temperature found in
(\ref{TSkelvin}),(\ref{TSkev}) for adiabatic shocks, hence the compression in isothermal
shocks may be orders of magnitude larger than the canonical
\newline
$(\gamma+1)/(\gamma-1)$
of adiabatic shocks.
This is a typical situation for formation of cool dense shells in interacting
supernovae.
The exact value of $T$ and of the compression depends on the details of the properties of plasma
with respect to heat conduction, but one should remember that those dense shells
may become unstable, and the exact numbers found in idealized accurate plane parallel
or spherically symmetric calculations may be not very useful.

\section{Numerical simulations of light curves}
% hydrodynamical instabilities in those events are discussed.

I describe some results of numerical simulations which take into account
radiation trapping effects in interacting supernovae.
For illustration I use the results from paper~[2]. % \cite{Sorokina2015}.

The simulations use presupernovae structures obtained either from
evolutionary codes or artificially constructed.
Anyway, the initial models have a fast moving part
which may be called ``ejecta''.
This part has mass $M_{\rm ej}$ and radius $R_{\rm ej}$.
$M_{\rm ej}$ can be much less than
the total mass of the collapsing core; it is just a convenient form of parametrization of models.

To make an interacting model the ejecta are surrounded
by a rather dense envelope, ``wind'',
with the mass $M_{\rm w}$ extended to the radius $R_{\rm w}$.
The outer radius of this envelope must be large $\sim 10^{16}$~cm or even larger
for extreme cases.
The envelope may have a power-law density distribution $\rho \propto r^{-p}$,
which simulates the wind that surrounds the exploding star.
For a steady wind, $p=2$,
but in the very last stages of the evolution of a presupernova star the wind may not be
steady and the parameter $p$ may be varied in the range between 1.5 and~3.5.
Another kind of envelope,
detached from the ejecta by a region of lower density, is also considered in our simulations.

The light curves are calculated for SNe exploding within these envelopes.
A shock wave forms at the border between the ejecta and the envelope.
The shock  very efficiently converts
the energy of the ordered motion of expanding gas to that of the chaotic
thermal motion of particles, which can be easily emitted.
As a result, one may expect to
obtain light curves powerful enough to explain at least a part of
superluminous SNe without an assumption of unusually high explosion energy.
The detailed computations support those expectations.

For type IIn SLSNe, that is narrow-line events, hydrogen rich envelopes are used.
For SLSN I (no hydrogen in spectra)
carbon-oxygen models with different C to O ratios or helium models
are typically employed.
The models may contain some amount of radioactive elements like \nifsx,
but it is not necessary in this class of simulations
since the effect of pure ejecta-CSM interaction is sufficient for
explaining majority of SLSNe with zero amount of \nifsx.

The synthetic light curves in [2] %\cite{Sorokina2015}
are calculated
using our multi-group radiation hydrodynamic code {\sc stella} in its
standard setup.
The code simulates spherically symmetric hydrodynamic flows coupled
with multi-group  radiative transfer.
The opacity routine takes into account electron scattering, free-free
and bound-free
processes.
Contribution of spectral lines (i.e. bound-bound processes) is treated
in approximation
of ``expansion'' opacity.

The explosions have been simulated as a ``thermal bomb'' with variable
energy $E_{\rm expl}$ of the order 2 - 4 foe (1 foe $=10^{51}$~ergs),
which is a bit larger than in a standard 1 foe supernova, but much lower
than invoked in hypernovae or in PISNe (pair-instability supernovae).

\begin{figure}[H]
\centering
\includegraphics[width=\linewidth]{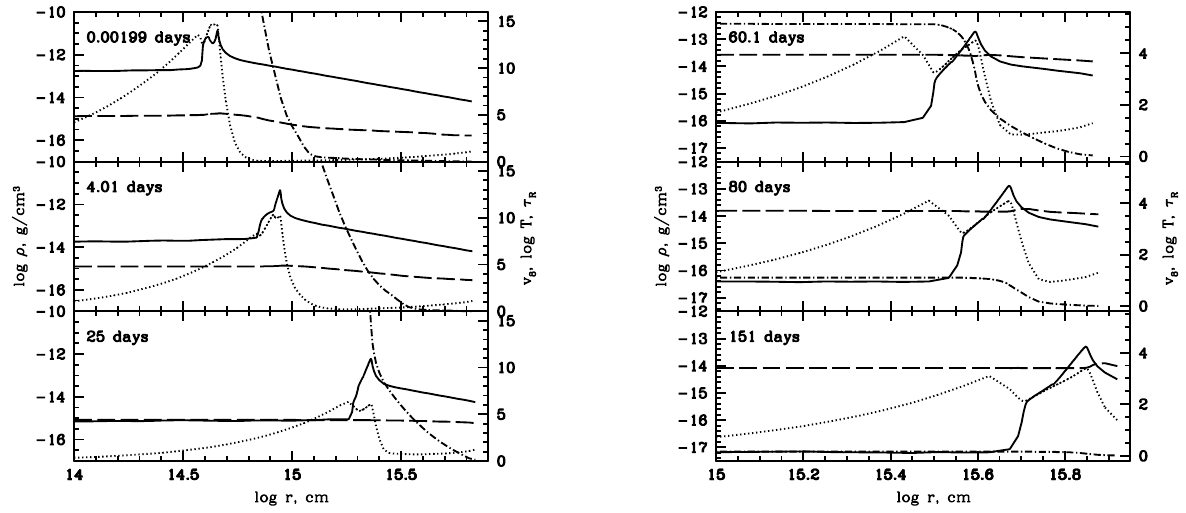}
\caption{\noindent\small
Evolution of radial profiles of the density ({\it solid lines}),
velocity (in $10^8$~cm~s$^{-1}$, {\it dots}), matter temperature ({\it dashes}),
and Rosseland optical depth ({\it dash-dots})
for one of the models [2]. % \cite{Sorokina2015}.
The scale for the density is on the left Y axis,
for all other quantities,  on the right Y axis.
Left panel: evolution of the hydrodynamical structure before maximum: very soon after the
explosion and at days 4 and 25.
Right panel: the same parameters, but after maximum: at days 60, 80, and 151.
Note that different scales for the axes are used on the left and right panels.
        }
\label{fig:hydroN0}
\end{figure}

Figure~\ref{fig:hydroN0}
shows how the profiles of density, velocity, temperature,
and Rosseland mean optical depth evolve along time for  one of the models. % N0 and B0.
The left panels correspond to the evolution before maximum of the light curve
(which happens on day 27 after the explosion for that model)
the right panels show the evolution after maximum.

At the very beginning,
the shock wave structure starts to form due to collision between the ejecta and the CSM.
Then the emission from the shock front heats the gas in the envelope, thus making it opaque,
and the photosphere moves to the outermost layers rather quickly.
When the photospheric radius reaches its maximum,
one can observe maximal emission from the supernova.

The speed of the growth of the photospheric radius depends on the mass of the envelope,
since more photons must be emitted from the shock to heat  larger mass envelopes.

% \end{multicols}

% \begin{multicols}{2}

Another parameter which impacts the initial growth of the photospheric radius
is the chemical composition of the envelope.
E.g., the light curve rises faster for a CO envelope than for a He one
as a lower temperature is needed to reach high opacity in a CO mixture.
This light curve behaviour can help set the composition for some observed SLSNe.

The plots on the right-hand side of Figure~\ref{fig:hydroN0} show  the stages
when the photosphere slowly moves back to the center,
and the envelope and the ejecta finally become fully transparent.
At the beginning of this post-maximum stage all gas in the envelope is already heated by the photons
which came from the shock region and diffused through the envelope to the outer edge,
and the whole system (ejecta and envelope) becomes almost isothermal.
The shock becomes weaker with time and emits fewer photons which can heat up the envelope,
so the temperature of the still unshocked envelope falls down.

The shocked material is gathered into a thin, dense layer % (see Fig.~\ref{shells}),
which finally contains almost all mass in the system.
Formation of this layer leads to numerical difficulties,
which significantly limit the time step of the calculation.
Another problem can also take place due to the thin layer formation:
a thin, dense shell with a very large radius would most probably be unstable and can
fragment into smaller lumps.
Then the problem would become essentially multi-dimensional.

On the velocity profiles, the multi-reflection structure forms from the very beginning.
It evolves very quickly to the standard two-shock (forward and reverse) picture.
This does not depend on the initial velocity profile in the envelope.
The interaction of the ejecta with the  envelope leads to similar
final velocity structures.
It looks like a self-similar behaviour analogous to the solution found by Nadyozhin and Chevalier
% \cite{Nadezhin1985ApSS} and \cite{Chevalier1982SelfSim}
but with radiation.

\begin{figure}[H]
\centering
\includegraphics[width=0.7\linewidth]{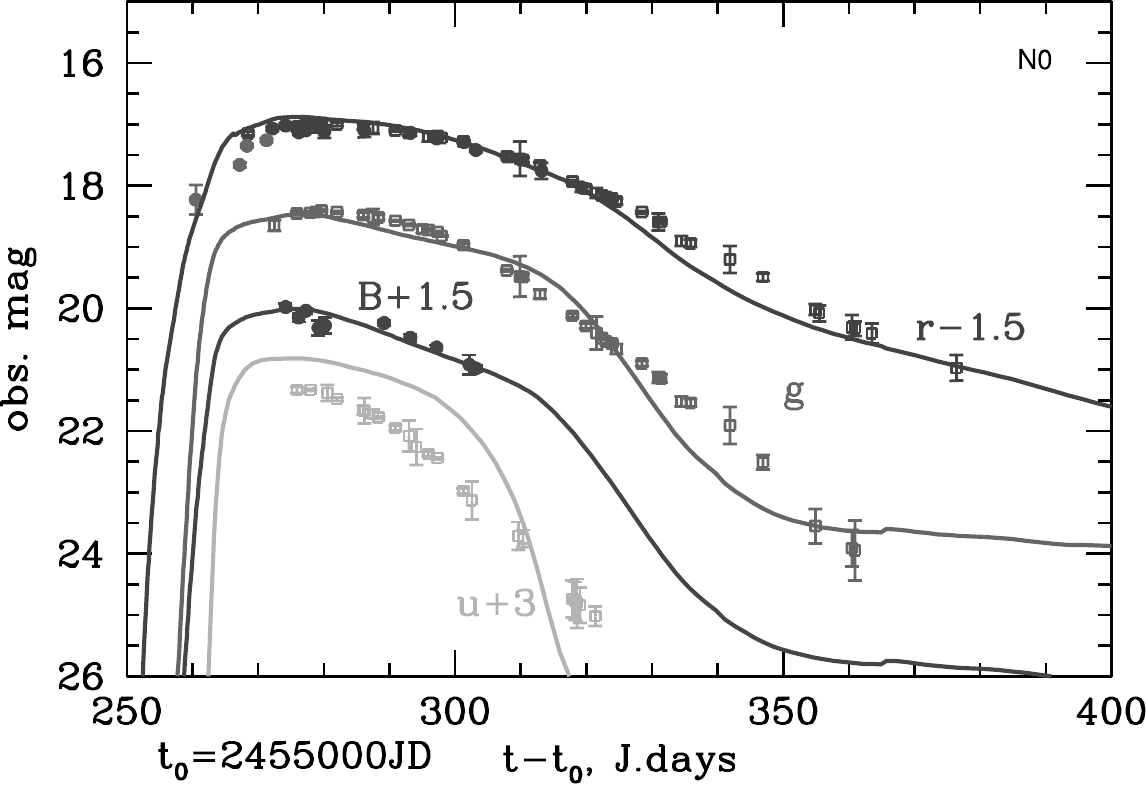}
\caption{\noindent\small Synthetic light curves for the model from Fig.~\ref{fig:hydroN0}, one of the best for SN~2010gx, in $r$, $g$, $B$, and $u$ filters compared with Pan-STARRS ($u$ and $g$ bands)
and PTF observations ($B$ and $r$).
        }
\label{pic:best10gx}
\end{figure}

Figure~\ref{pic:best10gx} demonstrates how the model with hydrodynamic evolution
shown in Figure~\ref{fig:hydroN0} reproduces multi-band observations
of the well studied Superluminous  SN~2010gx.

\section*{Acknowledgements}
The work on calculation of the SLSN light curves is supported
by a grant of Russian Science Foundation 14-12-00203.\\[0.2cm]

\baselineskip=10pt
\section*{References}

% \bibitem[Papadopoulos et al.(2015)]{2015MNRAS.449.1215P}
\small{\noindent
1. \textit{Papadopoulos, A.} \ DES13S2cmm: the first superluminous supernova from the Dark Energy Survey.\
\textit{A.~Papadopoulos, and 68 colleagues.} // Monthly Notices of the Royal Astronomical Society. 2015. V.~449, P.~1215---1227.
}

\small{\noindent
2. \textit{Sorokina,~E.I.}
Type I Superluminous Supernovae as Explosions inside Non-Hydrogen
  Circumstellar Envelopes}. \
\textit{E.I.~Sorokina, S.I.~Blinnikov, K.~{Nomoto}, R.~{Quimby}, A.~{Tolstov}}.\
// The Astrophysical Journal. 2016. V.~829, P.~17.
% \bibitem[Sorokina et al.(2016)]{2016ApJ...829...17S} Sorokina, E., Blinnikov, S., Nomoto, K., Quimby, R., Tolstov, A.\ 2016.\
% Type I Superluminous Supernovae as Explosions inside Non-hydrogen Circumstellar Envelopes.\
% The Astrophysical Journal 829, 17.

% \end{multicols}
\end{document}